\documentclass[12pt,a4paper]{article}
\usepackage{amssymb,amsfonts,amsmath,amsthm}
\usepackage{hyperref}
\author{Alexey.~A.~Magazev\thanks{E-mail: magazev@gmail.com}\\
{\small\sl Omsk State Technical University, Omsk, Russia}}

\title{Integrating Klein--Gordon--Fock equations in an external electromagnetic field on Lie groups}
\date{}

\newcommand{\eqdef}{\stackrel{\mathrm{def}}{=}}
\newcommand{\p}{\partial}
\newcommand{\g}{\mathfrak{g}}
\newcommand{\e}{\epsilon}
\renewcommand{\i}{\mathrm{i}}
\newcommand{\G}{\mathbf{G}}
\newcommand{\F}{\mathbf{F}}
\newcommand{\A}{\mathbf{A}}
\newcommand{\Ad}{\mathrm{Ad}}
\renewcommand{\O}{{\cal O}}
\newcommand{\ind}{\mathrm{ind}}
\newcommand{\pl}{\mathfrak{p}}
\renewcommand{\l}{\ell}
\newcommand{\D}{{\cal D}}

\newtheorem{statement}{Statement}
\newtheorem{theorem}{Theorem}
\newtheorem{corollary}{Corollary}

\begin{document}
\maketitle

\begin{abstract}
We investigate the structure of the Klein--Gordon--Fock equation symmetry algebra on pseudo-Riemannian
manifolds with motions in the presence of an external electromagnetic field. We show that in the case of an
invariant electromagnetic field tensor, this algebra is a one-dimensional central extension of the Lie algebra
of the group of motions. Based on the coadjoint orbit method and harmonic analysis on Lie groups, we
propose a method for integrating the Klein--Gordon--Fock equation in an external field on manifolds with
simply transitive group actions. We consider a nontrivial example on the four-dimensional group $E(2) \times \mathbb{R}$ in detail.
\end{abstract}

\section{Introduction}

The Klein-–Gordon--Fock (KGF) equation is the simplest relativistic equation describing the dynamics of massive spinless particles coupled to gauge--type fields. Despite its model character, this equation is rather often used with the goal of initially studying various quantum field effects: particle creation in external electromagnetic and gravitational fields, vacuum polarization, etc. \cite{GriMamMos80} -- \cite{Gal86}. In such cases, the problem of constructing the corresponding exact solutions seems crucial, when the standard approaches used in perturbation theory are limited or totally inapplicable.

A considerable number of works devoted to finding exact solutions of the KGF equation have currently accumulated (see, e.g., \cite{BagGit90}, \cite{BagBalGitShi01} and the references therein). We note that in the majority of them, classes of fields are considered that allow realizing a variable separation scheme in whose framework the KGF equation at least has a commutative symmetry algebra of at most second-order operators.

In \cite{ShaShi95}, a method for integrating linear partial differential equations was proposed using noncommutative
symmetry algebras. This method allowed a considerable broadening of the classification of external fields and pseudo-Riemannian manifolds admitting the existence of exact solutions of the KGF equation and served as a basis for developing original approaches to the study of quantum effects on homogeneous spaces \cite{BreShiMag11}. In particular, in the framework of the noncommutative integration method, all exactly solvable cases for the free KGF equation on four-dimensional Lie groups, i.e., on four-dimensional pseudo-Riemannian manifolds, admitting a simply transitive action of the group of motions were classified \cite{BarMikShi02}.

This paper is organized as follows. We first investigate the structure of the symmetry algebra of the KGF equation in an external electromagnetic field for an arbitrary pseudo-Riemannian manifold. We show that if the electromagnetic field 2-form is invariant under the action of the group of motions, then the free KGF equation symmetry operators associated with Killing vector fields can be extended to corresponding symmetry operators in the presence of a field. We prove a theorem stating that the operators in question form a one-dimensional central extension of the Lie algebra of the group of motions. This result in a sense generalizes the construction of gauge fields on homogeneous spaces proposed in \cite{KurShi08} in whose framework the symmetry--operator algebra is not deformed after a field is added. We also note that the defining equations for symmetry operators in external gauge fields were investigated by several authors \cite{Hol07}, \cite{Vis09}, but the corresponding symmetry algebra was practically not studied.

We then consider the situation where the group of motions acts on the manifold simply transitively in detail. In this case, the electromagnetic field is essentially identified with a certain 2-cocycle of the Lie algebra of the group of motions. To integrate the corresponding KGF equation in an external electromagnetic field, we use harmonic analysis on Lie groups including the construction of a special irreducible representation of a central extension of the Lie algebra of the group of motions ($\lambda$-representation) \cite{ShaShi95}. Using the $\lambda$-representation, we generalize the standard technique for the general solution of a Fourier series expansion, which reduces the initial problem to solving a reduced equation with fewer independent variables. We note that the solution in this case is constructed globally, i.e., the problem of "sewing"\ solutions from different charts of the initial manifold does not arise. This approach allows formulating a necessary and sufficient integrability condition expressed in terms of the so-called cohomological index, generalizing the notion of the usual Lie algebra index.

Finally, we consider an example of integrating the KGF equation in an invariant electromagnetic field on the four-dimensional Lie group that is a direct product of the group of Euclidean plane motions and the one-dimensional Abelian group $\mathbb{R}$ in detail.

\section{Algebra of KGF equation symmetry operators in an external electromagnetic field}

Let $(M, g)$ be a smooth connected manifold with a metric $g$. The metric structure is usually assumed to be Lorentzian in physical applications, but this is inessential in our case, and we therefore assume a pseudo-Riemannian metric. We consider the KGF equation
\begin{equation}
\label{eq:01}
\hat{H}\, \varphi \eqdef \left ( g^{ij} \nabla_i \nabla_j + m^2 \right ) \varphi = 0,
\end{equation}
in some coordinate chart $\{ x^i \}$ of the manifold $M$. Here, $g^{ij}$ are the contravariant components of metric tensor, $\nabla_i$ is the covariant derivative corresponding to the coordinate vector field $\p_i = \p/\p x^i$ (the connection is assumed to be consistent with the metric), and $m$ is a positive real parameter interpreted as the particle mass of the scalar field $\varphi(x)$. Here and hereafter, we assume summation over repeated indices.

We assume that the pseudo-Riemannian manifold $M$ admits an action of a connected group of motions $G$ defined by Killing vector fields $\xi_a = \xi_a^i(x) \p_i$, $a = 1, \dots , \dim G$. It is known that in this case, $\hat{\xi}_a = \xi_a^i \nabla_i$ are symmetry operators of Eq. \eqref{eq:01} and generators of the Lie algebra $\g$ of the group $G$:
\begin{equation}
\label{eq:02}
[\hat{H}, \hat{\xi}_a] = 0,\quad
[\hat{\xi}_a, \hat{\xi}_b] = C_{ab}^c\, \hat{\xi}_c,
\end{equation}
where $C_{ab}^c$ are structure constants of $\g$ in the basis defined by the vector fields $\xi_a$.

In addition, we assume that the manifold $M$ has a closed differential 2-form
\begin{equation*}
F = \sum \limits_{i < j} F_{ij}\, dx^i \wedge d x^j,\quad
dF = 0.
\end{equation*}
As follows from the Poincar\'e lemma, the closedness of the form implies $F = dA$, where $A = A_i\, d x^i$ is a
differential 1-form defined on $M$ at least locally. In what follows, we interpret the form $F$ as an external
electromagnetic field with the strength tensor $F_{ij}$ and the vector potential defined by the 1-form $A$.

It is known that the interaction of a charged scalar field with an electromagnetic field is taken into account by shifting covariant derivatives in the operator $\hat{H}$ by the corresponding vector potential components (see, e.g., \cite{GriMamMos80}): $\nabla_i \rightarrow \nabla_i^{(\e)} \eqdef \nabla_i - \i\e A_i$, where $\e$ is the electric charge of the scalar field particles and $\i^2 = -1$. After such a transformation, \eqref{eq:01} becomes the KGF equation in an external electromagnetic field
\begin{equation}
\label{eq:03}
\hat{H}^{(\e)}\, \varphi \eqdef \left ( g^{ij} \nabla_i^{(\e)} \nabla_j^{(\e)} + m^2 \right ) \varphi = 0.
\end{equation}

We investigate the question of symmetries of Eq. \eqref{eq:03}. To each Killing vector field $\xi_a$ we assign the
inhomogeneous differential operator $\hat{\xi}_a^{(\e)} = \xi_a^i \nabla_i^{(\e)} + \i\e \chi_a$, where the functions $\chi_a$ are defined by the condition
\begin{equation*}
[\hat{H}^{(\e)}, \hat{\xi}^{(\e)}_a] = 0.
\end{equation*}
Substituting the explicit form of the operators $\hat{\xi}^{(\e)}_a$ and $\hat{H}^{(\e)}$ here and taking into account that the vector fields $\xi_a$ are Killing, we obtain
\begin{equation*}
g^{ij} \left ( \p_i\, \chi_a + \xi_a^k F_{ki} \right ) \nabla_j^{(\e)} + \frac{1}{2}\, g^{ij} \nabla_i \left ( \p_j \, \chi_a + \xi_a^k F_{kj} \right ) = 0.
\end{equation*}
by a direct computation. It is easy to see that the obtained equality is equivalent to the condition
\begin{equation}
\label{eq:04}
\p_j \, \chi_a + \xi_a^k F_{kj} = 0,
\end{equation}
which can be represented invariantly in terms of differential forms as
\begin{equation}
\label{eq:05}
d \chi_a = - i_{\xi_a} F.
\end{equation}
Here, $i_{\xi} F \eqdef F_{ij}\xi^i d x^j$ is the inner product of the 2-form $F$ with the vector field $\xi$.

We regard \eqref{eq:05} as a system of equations for the unknown functions $\chi_a$. The system was obtained for
an arbitrary closed 2-form $F$ and is non-integrable in the general case.

\begin{statement}
Equality \eqref{eq:05} is equivalent to the requirement that the 2-form $F$ be invariant under the group $G$ action. In this case, the functions $\chi_a$ exist at least locally.
\end{statement}

\begin{proof}
Indeed, using the well-known differential geometry formula ${\cal L}_{\xi} = i_{\xi}\, d + d \, i_{\xi}$, we obtain
\begin{equation*}
d^2 \chi_a = - d \left ( i_{\xi_a} F \right ) = i_{\xi_a} d F - {\cal L}_{\xi_a} F = - {\cal L}_{\xi_a} F
\end{equation*}
for the 0-form $\chi_a$, where ${\cal L}_{\xi}$ is the Lie derivative along the vector field $\xi$. It is obvious that the right--hand side of the obtained equality is zero for all $a = 1, \dots, \dim G$ if and only if the group $G$ action preserves the 2-form $F$.

For a $G$-invariant 2-form $F$, the right-hand side of \eqref{eq:05} is a closed 1-form, and 0-forms $\chi_a$ hence exist at least locally.
\end{proof}

In what follows, we assume that the 2-form $F$ is $G$-invariant. The functions $\chi_a$ in this case are defined
up to additive constants and can be expressed in quadratures:
\begin{equation}
\label{eq:06}
\chi_a = - \int i_{\xi_a} F = - \int F_{ij}\, \xi_a^i\, dx^j.
\end{equation}

Using \eqref{eq:02}, we write the commutation relations satisfied by the inhomogeneous operators $\hat{\xi}^{(\e)}_a$:
\begin{equation*}
[\hat{\xi}^{(\e)}_a,\hat{\xi}^{(\e)}_b] = C_{ab}^c\, \hat{\xi}^{(\e)}_c + \i\e \left ( \hat{\xi}_a \chi_b - \hat{\xi}_b \chi_a - C_{ab}^c\, \chi_c - F(\xi_a, \xi_b) \right ).
\end{equation*}
It follows from \eqref{eq:05} that $\hat{\xi}_a \chi_b = F(\xi_a, \xi_b)$, whence we obtain
\begin{equation}
\label{eq:07}
[\hat{\xi}^{(\e)}_a,\hat{\xi}^{(\e)}_b] = C_{ab}^c\, \hat{\xi}^{(\e)}_c + \i\e\, \Omega_{ab},
\end{equation}
where we introduce the notation
\begin{equation}
\label{eq:08}
\Omega_{ab} = F(\xi_a, \xi_b) - C_{ab}^c\, \chi_c.
\end{equation}
It is easy to see that in the absence of an electromagnetic field or if $\e = 0$ (the case of a neutral field), equalities \eqref{eq:07} become commutation relations \eqref{eq:02}.

\begin{statement}
The functions $\Omega_{ab} \in C^{\infty}(M)$ defined by \eqref{eq:08} are constant on the manifold $M$ and have the properties
\begin{equation}
\label{eq:09}
\Omega_{ab} = - \Omega_{ba},
\end{equation}
\begin{equation}
\label{eq:10}
C_{ab}^d\, \Omega_{dc} + C_{bc}^d\, \Omega_{da} + C_{ca}^d\, \Omega_{db} = 0.
\end{equation}
\end{statement}

\begin{proof}
Using definition \eqref{eq:08} and condition \eqref{eq:05}, we obtain
\begin{multline*}
d \Omega_{ab} = d \left ( i_{\xi_b} i_{\xi_a} F - C_{ab}^c\, \chi_c \right ) = {\cal L}_{\xi_b} \left ( i_{\xi_a} F \right ) - i_{\xi_b} d \left ( i_{\xi_a} F \right ) - C_{ab}^c\, d \chi_c = \\
= - i_{[\xi_a, \xi_b]} F - i_{\xi_b} \left ( {\cal L}_{\xi_a} F - i_{\xi_a} dF \right ) + C_{ab}^c\, i_{\xi_c} F = 0
\end{multline*}
for the exterior differential of the 0-form $\Omega_{ab}$. Here, we use the formula ${\cal L}_{\xi}\, i_{\eta} F = - i_{[\xi,\eta]} F$ and also the closedness and $G$-invariance condition for the 2-form $F$.

Properties \eqref{eq:09} and \eqref{eq:10} are verified by a simple computation. In particular, \eqref{eq:10}) is a direct consequence of the form $F$ being closed and the Jacobi identity for structure constants $C_{ab}^c$ of the Lie algebra $\g$.
\end{proof}

We recall that a bilinear skew-symmetric function $\Omega: \g \times \g \rightarrow \mathbb{R}$ satisfying \eqref{eq:10} is called a \textit{2-cocycle} of the Lie algebra $\g$ with values in $\mathbb{R}$. We let $\mathbf{Z}^2(\g)$ denote the set of all 2-cocycles of $\g$. A one-dimensional \textit{central extension} $\tilde{\g} = \g \oplus_{\Omega} \mathbb{R}$ constructed by adding the one-dimensional center $\mathbb{R}$ to the algebra $\g$ corresponds to each $\Omega \in \mathbf{Z}^2(\g)$ \cite{Fuk84}. The commutation rule in the vector space $\tilde{\g}$ is
\begin{equation}
\label{eq:11}
[e_a, e_b] = C_{ab}^c\, e_c + \Omega_{ab}\, e_0,
\end{equation}
where $\{e_a\}$ is a basis in the algebra $\g$ and $e_0 \in \mathbb{R}$. Comparing commutation relations \eqref{eq:07} and \eqref{eq:11}, we can conclude that the span of the set $\hat{\xi}^{(e)}_0 = \i\e$, $\hat{\xi}^{(e)}_1, \dots, \hat{\xi}^{(e)}_{\dim \g}$ is a one-dimensional central extension of $\g$ corresponding to 2-cocycle \eqref{eq:08}.

We summarize the obtained results in the following theorem.

\begin{theorem}
\label{th:01}
Let $(M,g)$ be a pseudo-Riemannian manifold with an action of the motion group $G$ defined by the Killing vector fields $\xi_a = \xi_a^i \p_i$, and let $F$ be an invariant closed 2-form on $M$. Then the KGF equation in an external electromagnetic field admits the symmetry operator Lie algebra $\tilde{\g}$ whose basis elements can be chosen in the form
\begin{equation*}
\hat{\xi}^{(\e)}_0 = \i\e,\quad
\hat{\xi}^{(\e)}_a = \xi_a^i \nabla^{(\e)}_i + \i\e \chi_a,\quad
a = 1, \dots, \dim \g,
\end{equation*}
where the functions $\chi_a \in C^{\infty}(M)$ are defined by \eqref{eq:06}. The Lie algebra $\tilde{\g}$ is a one-dimensional central extension of the Lie algebra $\g$ of the motion group $G$ defined by 2-cocycle \eqref{eq:08}.
\end{theorem}

We note that \eqref{eq:09} and \eqref{eq:10} are satisfied, for example, by the bilinear functions $\Omega_{ab} = C_{ab}^c\, \lambda_c$, where $\lambda_c$ is a constant. Such 2-cocycles are said to be \textit{trivial} or are called \textit{2-coboundaries}. The set of all $\g$ 2-coboundaries is denoted by $\mathbf{B}^2(\g)$ in what follows.

It is easy to see that the central extension $\tilde{\g}$ corresponding to a trivial cocycle can be decomposed into
a direct sum $\tilde{\g} = s(\g) \oplus \mathbb{R}$, where the map $s$ is a Lie algebra isomorphism. Indeed, if $\Omega_{ab} = C_{ab}^c\, \lambda_c$, then commutation relations \eqref{eq:07} can be represented as
\begin{equation*}
[\hat{\xi}^{(\e)}_a,\hat{\xi}^{(\e)}_b] = C_{ab}^c \left  ( \hat{\xi}^{(\e)}_c + \i\e\lambda_c \right ).
\end{equation*}
Because the $\chi_a$ are given up to additive constants, the isomorphism $s$ is established by the variable change
$\chi_a \rightarrow \chi_a + \lambda_a$. In such cases, it is also said that the central extension $\tilde{\g}$ of the algebra $\g$ is \textit{splitting} \cite{GotGro78}.

Let $F_1$ and $F_2$ be invariant closed 2-forms on $M$ describing two different electromagnetic field configurations.
We let $\Omega_1$ and $\Omega_2$ denote the corresponding 2-cocycles and consider central extensions $\tilde{\g}_1$ and $\tilde{\g}_2$ corresponding to these cocycles. The extensions $\tilde{\g}_1$ and $\tilde{\g}_2$ of $\g$ are said to be equivalent if $\Omega_2 - \Omega_1 \in \mathbf{B}^2(\g)$, i.e., the 2-cocycles $\Omega_1$ and $\Omega_2$ differ by a coboundary. It is easy to see that equivalent central extensions are isomorphic Lie algebras. Therefore, there is interest in considering only non-equivalent central extensions that are in one-to-one correspondence with elements of the quotient space $\mathbf{H}^2(\g) \eqdef \mathbf{Z}^2(\g) / \mathbf{B}^2(\g)$. The space $\mathbf{H}^2(\g)$ equipped with the natural Abelian group structure is called the \textit{2-cohomology group} of the Lie algebra $\g$ \cite{Fuk84}, \cite{GotGro78}.

\section{The KGF equation in an external electromagnetic field on Lie groups}

We assume that a connected group $G$ acts on a pseudo-Riemannian manifold $(M,g)$ \textit{simply transitively}. This means that for any two points $x, x' \in M$, there exists precisely one shift in $G$ transforming $x$ to $x'$. There is thus a smooth one-to-one correspondence between the points of $M$ and the elements of $G$. In what follows, we build all the necessary constructions on the Lie group $G$ itself, keeping the diffeomorphism $M \simeq G$ in mind. For definiteness, we assume that the group acts on itself by right shifts.

It is known that the set of right-invariant pseudo-Riemannian metrics on $G$ is in a one-to-one correspondence with the set of bilinear symmetric nondegenerate forms on the Lie algebra $\g \simeq T^*_e G$. We fix a basis $\{e_a\}$ in $g$ and choose the basis $\{e^a\}$ in the dual space $\g^*$ such that $\langle e^a, e_b \rangle = \delta^a_b$, $a, b = 1, \dots, \dim \g$.
In this case, the above correspondence is given by the rule
\begin{equation}
\label{eq:12}
g_{ij}\, dx^i \otimes dx^j = \G_{ab}\, \sigma^a \otimes \sigma^b,
\end{equation}
where $\sigma^a(x) = (R_{x^{-1}})^*\, e^a$ are right-invariant 1-forms on $G$ corresponding to the basis elements $e^a \in \g^*$ and $\G = g|_{x = e}$ is a nondegenerate symmetric form on the tangent space at the group unity. It is easy to see that left-invariant vector fields $\xi_a(x) = (L_x)_*\, e_a$, generators of the right regular representation of $G$, are Killing
vector fields for metric \eqref{eq:12}.

The Laplace--Beltrami operator constructed using metric \eqref{eq:12} can be represented as a second-order operator polynomial in the basis right-invariant vector fields $\eta_a(x) = (R_x)_*\, e_a$:
\begin{equation}
\label{eq:13}
\hat{H} = g^{ij} \nabla_i \nabla_j = \G^{ab} \left ( \eta_a + C_a \right ) \eta_b,
\end{equation}
where $C_a = C_{ab}^b$ and $\G^{ab} \G_{bc} = \delta^a_c$. Because left- and right-invariant vector fields on Lie groups commute, it is obvious that $[\hat{H}, \hat{\xi}_a] = 0$ follows from \eqref{eq:13}. We also note that if the group $G$ is unimodular, then $C_a = 0$, and expression \eqref{eq:13} for the Laplace--Beltrami operator hence simplifies to $\hat{H} = \G^{ab} \eta_a \eta_b$.

We assume that an external electromagnetic field is given on the manifold of the Lie group $G$ by means of a closed 2-form $F$ invariant under right shifts. To the form $F$, we can bijectively assign the bilinear skew-symmetric form $F = (1/2)\, \F_{ab}\, e^a \wedge e^b$ defined on $\g$ such that
\begin{equation}
\label{eq:14}
F_{ij}\, dx^i \wedge dx^j = \F_{ab}\, \sigma^a \wedge \sigma^b.
\end{equation}
It is easy to show that 2-form \eqref{eq:14} being closed is equivalent to the condition
\begin{equation}
\label{eq:15}
C_{ab}^d\, \F_{dc} + C_{bc}^d\, \F_{da} + C_{ca}^d\, \F_{db} = 0
\end{equation}
on the bilinear form $\F$. As noted in the preceding section, condition \eqref{eq:15} means that the form $\F$ is a 2-cocycle of $\g$ with values in $\mathbb{R}$. Consequently, the set of invariant closed 2-forms on $G$ is in one-to-one correspondence with the set $\mathbf{Z}^2(\g)$.

We note that formulas \eqref{eq:12} and \eqref{eq:14} are \textit{tetradic decompositions} of the tensors $g_{ij}$ and $F_{ij}$ over the right-invariant tetrad of 1-forms $\{\sigma^a\}$.

A construction of a vector potential corresponding to the strength tensor of a right-invariant electromagnetic field on a Lie group was given in \cite{MagShiYur08}. For the completeness of our exposition, we here present the basic results concerning this question.

We represent the vector potential 1-form $A = A_i\, dx^i$ as a decomposition over the tetradic basis of right-invariant 1-forms, $A = \A_a\,\sigma^a$, where the components $\A_a$ are, in general, functions in $C^{\infty}(G)$. Obviously, the transition from the $\A_a$ to the tensor components $A_i$ is bijective: $A_i = \sigma^a_i\, \A_a$, $\A_a = \eta^i_a \A_i$. It is easy to show that solving the equation $dA = F$ in this case reduces to solving the equivalent problem
\begin{equation}
\label{eq:16}
\eta_a \A_b - \eta_b \A_a + C_{ab}^c\, \A_c = \F_{ab}
\end{equation}
for the unknown functions $\A_a$.

System of differential equations \eqref{eq:16} is (locally) integrable because the constant functions $\F_{ab}$ satisfy \eqref{eq:15}. The general solution can be represented as $\A_a = \A_a' + \eta_a S$, where $S \in C^{\infty}(G)$ is an arbitrary function and $\A'_a$ α is some particular solution of system \eqref{eq:16}. The freedom in choosing the solution is in fact a manifestation of the gauge invariance of the equation $dA = F$.

The simplest particular solution of system \eqref{eq:16} is constructed for a trivial 2-cocycle. Indeed, the following statement holds.

\begin{statement}
\label{st:03}
Let $\F \in \mathbf{B}^2(\g)$; hence, $\F_{ab} = C^c_{ab}\, \lambda_c$ for some dual vector in $\lambda \in \g^*$. Then the
particular solution of system \eqref{eq:16} can be chosen in the form $\A'_a = \lambda_a$, $a = 1, \dots, \dim \g$.
\end{statement}

As a consequence, we note that the particular solution in Proposition \ref{st:03} corresponds to a vector potential $A$ invariant under right shifts on $G$ (this is not the case in general!).

To formulate a corresponding result in the general case, we need certain additional constructions. Let $\tilde{\g} = \g \oplus_{\F} \mathbb{R}$ be a one-dimensional central extension of $\g$ corresponding to the 2-cocycle $\F$. Let $\tilde{G}$ denote the connected simply connected Lie group whose Lie algebra is isomorphic to $\tilde{\g}$. In this case, a connected one-dimensional subgroup $Z$ in the center of $\tilde{G}$ corresponds to the one-dimensional ideal $\mathbb{R} \subset \tilde{\g}$. Obviously, the quotient group $G' = \tilde{G}/Z$ is locally isomorphic to $G$, and therefore $G = G'/\Gamma$, where $\Gamma$ is a discrete normal subgroup in $G'$ belonging to the center of $\tilde{G}$. We hence have $G \simeq \tilde{G}/(Z \times \Gamma)$.

\begin{statement}
In a trivialization $V \subset G$ of the principal bundle $G \simeq \tilde{G}/(Z \times \Gamma)$, a particular solution of system \eqref{eq:16} can be found by the formula 
\begin{equation}
\label{eq:17}
\A'_a = - U^{-1} \tilde{\eta}_a U \big|_{V},\quad
a = 1, \dots, \dim \g,
\end{equation}
where $U : Z \rightarrow GL(1; \mathbb{R})$ is a one-dimensional real representation of the group $Z$ and $\tilde{\eta}_a$ is a right-invariant
field on the Lie group $\tilde{G}$ corresponding to the basis element $e_a \in \g$.
\end{statement}

A proof of the statement is given in \cite{MagShiYur08}. As a note, we emphasize that the proposed particular solution is local. Indeed, as some simplest examples show, a global solution of system \eqref{eq:16} can in general not exist on the whole group manifold (an obstruction to this is the presence of a nontrivial second cohomology group $H^2(G)$). For example, a closed volume 2-form on the two-dimensional group $S^1 \times S^1$, as is known, is not exact.

In the KGF equation, we switch on a right-invariant electromagnetic field defined using 2-form \eqref{eq:14}. We assume that the metric is right-invariant, i.e., the corresponding squared length element has form \eqref{eq:12}. In this case, the operation of covariant derivative shift in Laplace--Beltrami operator \eqref{eq:13} is equivalent to the variable change $\eta_a \rightarrow \eta^{(\e)}_a \eqdef \eta_a - \i\e \A_a$, $\A_a = \A'_a + \eta_a S$, where $\A_a'$ is defined by \eqref{eq:17} and $S \in C^{\infty}(G)$ is an arbitrary function on the group:
\begin{equation}
\label{eq:18}
\hat{H}^{(\e)} = \G^{ab}\, \left ( \eta^{(\e)}_a + C_a \right ) \eta^{(\e)}_b.
\end{equation}

In general, the constructed operator $\hat{H}^{(\e)}$ does not commute with the left-invariant field $\xi_a$. It can be shown that the equality $[\hat{H}^{(\e)}, \xi_a] = 0$ holds if and only if the corresponding 2-cocycle $\F$ is trivial, i.e.,
$\F \in \mathbf{B}^2(\g)$ (see Statement \ref{st:03}). It follows from Theorem \ref{th:01} that in the general case, operators commuting with \eqref{eq:18} have the form
\begin{equation}
\label{eq:19}
\xi^{(\e)}_a = \xi_a + \i\e \left ( \chi_a - (\Ad_x)^b_a \, \A_b \right ),
\end{equation}
where $\chi_a$ are defined by formula \eqref{eq:06} and $\Ad_x \eqdef (R_{x^{-1}})_* (L_x)_*$ is the coadjoint representation
matrix for the group $G$. The linear span of the operators $\xi^{(\e)}_0 = \i\e$ and \eqref{eq:19} forms the Lie algebra $\tilde{\g}'$, which is a one-dimensional central extension of $\g$ defined using 2-cocycle \eqref{eq:08}.

\begin{statement}
\label{st:05}
The Lie algebras $\tilde{\g}$ and $\tilde{\g}'$ are isomorphic.
\end{statement}

\begin{proof}
We recall that two central extensions of a Lie algebra $\g$ are isomorphic if and only if the corresponding 2-cocycles differ by a trivial cocycle, i.e., belong to the same cohomology class \cite{Fuk84}.

The value of the right-invariant 2-form $F$ on the left-invariant vector fields $\xi_a = (L_x)_*\, e_a$ and $\xi_b = (L_x)_*\, e_b$ is
\begin{equation}
\label{eq:20}
F(\xi_a, \xi_b) = \F\left ( (R_{x^{-1}})_*\, (L_x)_*\, e_a, (R_{x^{-1}})_*\, (L_x)_*\, e_b \right ) = \F(\Ad_x e_a, \Ad_x e_b).
\end{equation}
The equality
\begin{equation}
\label{eq:21}
\F(\Ad_x e_a, \Ad_x e_b) = \F(e_a, e_b) + C_{ab}^c\, \Theta_c(x)
\end{equation}
holds, where $\Theta_c \in C^{\infty}(G)$ is some set of functions. Indeed, an infinitesimal analogue of this relation is formula \eqref{eq:15}, which is also a sufficient condition for the validity of Eq. \eqref{eq:21} because the group $G$ is
connected (also see p. 175 in \cite{Kir02}). Taking \eqref{eq:20} and \eqref{eq:21} into account and using formula \eqref{eq:08}, we obtain
\begin{equation*}
\Omega_{ab} - \F_{ab} = C_{ab}^c \left ( \Theta_c - \chi_c \right )
\end{equation*}
for the difference of the 2-cocycles of the Lie algebras $\tilde{\g}$ and $\tilde{\g}'$. The left-hand side of this equality is a constant, and we can therefore pick a value of the function in the right-hand side at any point, for example, at the unity of the group $G$. From this, we directly obtain $\Omega - \F \in \mathbf{B}^2(\g)$.
\end{proof}

\section{Integration method}

We next consider the question of the integrability of the KGF equation in an external electromagnetic field on a group $G$ and also give an algorithm for constructing its general solution. In this context, we say that Eq. \eqref{eq:03} is integrable if finding its solution reduces to solving an ordinary differential equation and computing quadratures.

We note that operator \eqref{eq:18} has a symmetry operator algebra of form \eqref{eq:19} by construction. Generally speaking, the above problem has no other symmetries, and the most efficient way of constructing exact solutions of Eq. \eqref{eq:03} is therefore the method for noncommutative integration of linear differential equations based on the Kirillov orbit method \cite{Kir02}, \cite{Kir76} and harmonic analysis on Lie groups. Unfortunately, the size of a journal paper does not allow expressing all the statements and theorems of this method with sufficient argumentation, and we therefore only recall the necessary constructions, adapting them to solve our problem (see \cite{ShaShi95}, \cite{BreShiMag11}, \cite{GonShi09} for the details of the method).

Let $\tilde{\g}$ be a one-dimensional central extension of the Lie algebra $\g$ constructed using the 2-cocycle $\F \in \mathbf{Z}^2(\g)$ and $\tilde{G}$ be the corresponding connected simply connected Lie group. In what follows, we identify the dual space $\tilde{\g}^*$ with the direct sum $\g^* \oplus \mathbb{R}$. The group $\tilde{G}$ acts by the coadjoint representation in the dual space $\tilde{\g}^*$, fibering it into coadjoint orbits. A closed invariant nondegenerate 2-form (the Kirillov form) is defined on each orbit, equipping the orbit with a symplectic manifold structure. In particular, all coadjoint orbits are even-dimensional.

We let $\O_{\lambda,\e}$ denote the orbit passing through the linear functional $\lambda \oplus \e \in \tilde{\g}^*$, where $\lambda \in \g^*$ and $\e \in \mathbb{R}$ is the particle charge of the scalar field. If the above element is generic, then the orbit $\O_{\lambda,\e}$ is said to be nondegenerate, and its dimension is
\begin{equation*}
\dim \O_{\lambda,\e} = \dim \g -\ind_{[\F]}\, \g 
\end{equation*}
where $\ind_{[\F]}\, \g$ is our notation for the \textit{cohomological index} of the Lie algebra $\g$ of a class $[\F] \in \mathbf{H}^2(\g)$, introduced in \cite{MagShiYur08}
\begin{equation*}
\ind_{[\F]}\, \g \eqdef \sup \limits_{\F' \in [\F]} \dim \left ( \ker \F' \right ).
\end{equation*}
We note that for the goals of harmonic analysis, we must consider only nondegenerate orbits, and we therefore restrict ourself to just this case in what follows.

We recall that for an arbitrary Lie algebra $\g$, the subalgebra $\pl \in \g^{\mathbb{C}}$ is called a \textit{polarization} of a linear functional $\lambda \in \g^*$ if it is subordinate to the element $\lambda$ and its codimension equals half the dimension of the orbit $\O_{\lambda} \subset \g^*$,
\begin{equation*}
\langle \lambda, [\pl, \pl] \rangle = 0,\quad
\dim \pl = \dim \g - \frac{1}{2}\, \dim O_{\lambda}.
\end{equation*}
It was established in \cite{Dix77} that for any generic functional in the space $\g^*$, there exists a polarization.

Let $\tilde{\pl}$ denote a polarization of the linear functional $\lambda \oplus \e \in \tilde{\g}^*$. We have the inclusion $\mathbb{R} \in \tilde{\pl}$, and an arbitrary polarization in $\tilde{\g}^{\mathbb{C}}$ therefore decomposes into the direct sum $\tilde{\pl} = \pl^{\e} \oplus \mathbb{R}$, where $\pl^{\e} \subset \g^{\mathbb{C}}$. It can be shown that $\pl^{\e}$ is a subalgebra in $\g^{\mathbb{C}}$, which is a polarization of $\lambda$ if $\e = 0$.

It follows from the results in the preceding section that the correspondence $e_a \rightarrow - \eta^{(\e)}_a$ yields a
representation of $\tilde{\g}$ in the space of complex-valued functions on the group $G$. Here, $\eta^{(\e)}_a = \eta_a - \i\e\A_a$, where $\eta_a$ is the right-invariant vector field corresponding to the basis vector $e_a$ and the functions $\A_a$ are defined by \eqref{eq:16}. We extend this representation by linearity to a representation of the complex span $\tilde{\g}^{\mathbb{C}}$ and define the functional subspaces $L(G,\pl^{(\e)})$ of solutions of the system
\begin{equation*}
X^a \left ( \eta^{(\e)}_a - \i\lambda_a \right ) \varphi(x) = 0,\quad
X^a e_a \in \pl^{\e}.
\end{equation*}
There is a local isomorphism $\iota : L(G, \pl^{\e}) \rightarrow L(Q, \pl^{\e})$ (see \cite{Kir76}, pp. 224--226 in the Russian version), where $L(Q, \pl^{\e})$ is the space of complex-valued functions on the mixed manifold\footnote{The space $Q$ is called a \textit{mixed} type-$(k, l)$ manifold if it is locally isomorphic to a zero neighborhood in the space $\mathbb{R}^k \oplus \mathbb{C}^l$ and the corresponding transition functions are infinitely differentiable in the real coordinates in $\mathbb{R}^k$ and holomorphic in the complex coordinates in $\mathbb{C}^l$.} $Q$, whose dimension equals half the dimension of the orbit $\O_{\lambda,\e}$. Because $L(G, \pl^{\e})$ is invariant under right shifts, the linear operators $\l_a$ acting in the space $L(Q, \pl^{\e})$ are well-defined:
\begin{equation*}
\l_a = \iota \circ \xi^{(\e)}_a \circ \iota^{-1},
\end{equation*}
where $\xi^{(\e)}_a$ are given by \eqref{eq:19}. The operators $\ell_a$ realize the irreducible representation of $\tilde{\g}$ in $L(Q, \pl^{(\e)})$ that is called the $\lambda$-\textit{representation}. In the local coordinates of the manifold $Q$, the $\lambda$-representation operators are inhomogenous first-order differential operators depending on $\dim Q = (\dim \g - \ind_{[\F]}\, \g )/2$ independent variables.

On the mixed manifold $Q$ for further purposes, we introduce the measure $d\mu(q)$ and the inner product
\begin{equation}
\label{eq:22}
(\psi_1, \psi_2)_Q = \int \limits_Q \overline{\psi_1(q)} \psi_2(q) d\mu(q),
\end{equation}
with respect to which the $\lambda$-representation operators are anti-Hermitian\footnote{For this, we must introduce the "quantum shift"\ $\lambda \rightarrow \lambda + \i \vartheta$ in $\l_a$, where $\vartheta \in \g^*$ is chosen from the reality condition for the Casimir operator spectrum in the $\lambda$-representation.}: $\l_a^{+} = - \l_a$.

We consider the family of generalized functions $\D^{\lambda,\e}_{q,\bar{q}'}(x)$ on $G$ satisfying the system of equations
\begin{equation}
\label{eq:23}
\left ( \eta_a^{(\e)} - \l_a(q,\lambda)\right ) \D^{\lambda,\e}_{q,\bar{q}'}(x) = 0,\quad
\left ( \xi_a^{(\e)} + \overline{\l_a(q',\lambda)} \right ) \D^{\lambda,\e}_{q,\bar{q}'}(x) = 0.
\end{equation}
The equality $\xi^{(\e)}_a(e) - \eta^{(\e)}_a(e) = \i\e \chi_a(e)$ follows from \eqref{eq:19}. We use freedom in choosing the additive constant in \eqref{eq:06} such that the condition $\chi_a(e) = 0$ is satisfied. With the normalization
\begin{equation}
\label{eq:24}
\D^{\lambda,\e}_{q,\bar{q}'}(e) = \delta(q,\bar{q}')
\end{equation}
taken into account, the solution of system \eqref{eq:23} is then found uniquely. Here, $\delta(q,\bar{q}')$ is a delta function on the manifold $Q$ defined with respect to inner product \eqref{eq:22}.

The orthogonality and completeness of the functions $\D^{\lambda,\e}_{q,\bar{q}'}(x)$,
\begin{equation}
\label{eq:25}
\int \D^{\lambda,\e}_{q,\bar{q}'}(x) \overline{\D^{\tilde{\lambda},\e}_{\tilde{q},\bar{\tilde{q}}'}(x)}\, d\mu(x) = \delta(\lambda,\tilde{\lambda}) \delta(q,\bar{\tilde{q}}) \delta(q', \bar{\tilde{q}}');
\end{equation}
\begin{equation}
\label{eq:26}
\int \D^{\lambda,\e}_{q,\bar{q}'}(x) \overline{\D^{\lambda,\e}_{q,\bar{q}'}(\tilde{x})}\, d\mu(q) d\mu(q') d\mu(\lambda) = \delta(x, \tilde{x})
\end{equation}
allow defining the direct and inverse Fourier transformations:
\begin{equation}
\label{eq:27}
\psi_{\lambda,\e}(q, \bar{q}') = \int \D^{\lambda,\e}_{q,\bar{q}'}(x) \varphi(x) d\mu(x),
\end{equation}
\begin{equation}
\label{eq:28}
\varphi(x) = \int \overline{\D^{\lambda,\e}_{q,\bar{q}'}(x)} \psi_{\lambda,\e}(q, \bar{q}') d\mu(\lambda) d\mu(q) d\mu(q').
\end{equation}
It is crucial for the subsequent arguments that the actions of the operators $\eta_a^{(\e)}$ and $\xi^{(\e)}_a$ after transformation \eqref{eq:27} are mapped to the action of the corresponding $\lambda$-representation operators, and vice versa:
\begin{equation*}
\{ \varphi(x) \Leftrightarrow \psi_{\lambda,\e}(q, \bar{q}') \} \Longrightarrow \{ \eta^{(\e)}_a \varphi(x) \Leftrightarrow \l_a(q,\lambda) \psi_{\lambda,\e}(q,\bar{q}'), \xi^{(\e)}_a \varphi(x) \Leftrightarrow \overline{\l_a(q',\lambda)} \psi_{\lambda,\e}(q,\bar{q}')\}.
\end{equation*}

The inhomogenous operators $\eta^{(\e)}_a$ and $\xi^{(\e)}_a$ depend on $\dim \g$ variables, while the operators $\l_a(q,\lambda)$ depend on $(\dim \g - \ind_{[\F]}\, \g)/2$ variables. In particular, this allows using direct and inverse Fourier transformations \eqref{eq:27} and \eqref{eq:28} to reduce the differential operators that are polynomial combinations of the operators $\eta^{(\e)}_a$ to operators with fewer variables.

We apply the above scheme to the KGF equation in an external electromagnetic field with right-invariant 2-form \eqref{eq:14}. The operator $\hat{H}^{(\e)}$ is represented by expression \eqref{eq:18}. Therefore, Eq. \eqref{eq:03} after reduction becomes an equation for an unknown function $\psi_{\lambda,\e}(q,\bar{q}')$:
\begin{equation}
\label{eq:29}
\G^{ab} \left ( \l_a(q,\lambda) + C_a \right ) \l_b(q,\lambda) \psi_{\lambda,\e}(q,\bar{q}') = - m^2 \psi_{\lambda,\e}(q,\bar{q}').
\end{equation}
We note that the variables $q'$ enter the equation as parameters.

If we succeed in constructing the general solution of \eqref{eq:29}, then we can use quadrature \eqref{eq:28} to obtain the general solution of \eqref{eq:03}. In particular, if $\dim Q = 1$, then \eqref{eq:29} is an ordinary differential equation, and
Eq. \eqref{eq:03} is consequently integrable in this case. We have thus proved the following theorem.

\begin{theorem}
\label{th:02}
Let the electromagnetic field on the group $G$ be given using right-invariant closed 2-form \eqref{eq:14} corresponding to the 2-cocycle $\F \in \mathbf{Z}^2(\g)$. Then Eq. \eqref{eq:03} for an arbitrary right-invariant metric on $G$ reduces to Eq. \eqref{eq:29} with $\dim Q = (\dim \g - \ind_{[\F]}\, \g)/2$ independent variables. In particular, Eq. \eqref{eq:03} is integrable if
\begin{equation*}
\frac{1}{2} \left ( \dim \g - \ind_{[\F]}\, \g \right ) \leq 1.
\end{equation*}
\end{theorem}

For semisimple Lie algebras, the second Whitehead lemma stating that $\mathbf{H}^2(\g) = 0$ is applicable (see, e.g., \cite{GotGro78}). This means that any 2-cocycle $\F$ of a semisimple Lie algebra $\g$ is trivial. It was shown in \cite{MagShiYur08} that in this case, $\ind_{[\F]}\, \g = \ind\, \g$, i.e., the cohomological index coincides with the usual index of the Lie algebra $\g$. It follows that the integrability of free KGF equation \eqref{eq:01} is "preserved" for semisimple Lie groups even when a right-invariant electromagnetic field is switched on. Obviously, a Lie group with the
zero 2-cohomology group $\mathbf{H}^2(\g)$ has this property.

\begin{corollary}
Let $G$ be a semisimple Lie group. Then Eq. \eqref{eq:03} for an arbitrary right-invariant metric in an external right-invariant electromagnetic field is integrable if the free KGF equation in the absence of the field is integrable, i.e., in the case where
\begin{equation*}
\frac{1}{2} \left ( \dim \g - \ind\, \g \right ) \leq 1.
\end{equation*}
\end{corollary}

\section{Integration of the KGF equation in an external electromagnetic field on the group $E(2) \times \mathbb{R}$}

We illustrate the results presented above with a concrete example. We consider the four-dimensional connected Lie group $G = E(2) \times \mathbb{R}$, where $E(2)$ is the group of motions of a two-dimensional Euclidean plane. The Lie algebra $\g = \mathfrak{e}(2)\oplus \mathbb{R}$ of $G$ is defined by the commutation relations
\begin{equation*}
[e_1, e_2] = 0,\quad
[e_1, e_3] = e_2,\quad
[e_2, e_3] = - e_1,\quad
[e_a, e_4] = 0, \quad
a = 1, 2, 3;
\end{equation*}
where $\{e_1, e_2, e_3\}$ is a basis in the Lie algebra $\mathfrak{e}(2)$ of $E(2)$ and $\{e_4\}$ is a basis in the one-dimensional center $\mathbb{R}\simeq \g/\mathfrak{e}(2)$.

The algebra $\g$ admits the four-dimensional 2-cocycle space $\mathbf{Z}^2(\g)$, an arbitrary of which element can be represented as
\begin{equation}
\label{eq:30}
\F = \mu_1\, e^1 \wedge e^2 + \mu_2\, e^3 \wedge e^4 + \mu_3\, e^1 \wedge e^3 + \mu_4\, e^2 \wedge e^3,
\end{equation}
where $\{e^a\}$ is the dual basis in $\g^*$. The inclusion of the trivial cocycle subspace $\mathbf{B}^2(\g)$ in $\mathbf{Z}^2(\g)$ is described by the equalities $\mu_1 = \mu_2 = 0$. Hence, $\dim \mathbf{H}^2(\g) = 2$.

We introduce local coordinates of the second kind in a neighborhood $V \subset G$ of the group unity: $g(x) = e^{x_1 e_1} e^{x_2 e_2} e^{x_3 e_3} e^{x_4 e_4}$ (the range of each coordinate except $x_3 \in (−\pi, \pi)$ is the entire set $\mathbb{R}$). In the
chosen coordinates, the left- and right-invariant basis vector fields on $G$ become
\begin{equation}
\label{eq:31a}
\xi_1 = \cos x_3\, \p_1 - \sin x_3\, \p_2,\quad
\xi_2 = \sin x_3\, \p_1 + \cos x_3\, \p_2,\quad
\xi_3 = \p_3,\quad
\xi_4 = \p_4,
\end{equation}
\begin{equation*}
\label{eq:31b}
\eta_1 = \p_1,\quad
\eta_2 = \p_2,\quad
\eta_3 = x_2\, \p_1 - x_1\, \p_2 + \p_3,\quad
\eta_4 = \p_4.
\end{equation*}
The group $G$ is unimodular, and as a concrete measure on $G$, we can chose
\begin{equation*}
d\mu(x) = \frac{1}{2\pi}\, dx_1 dx_2 dx_3 dx_4.
\end{equation*}

We write the coordinate representation of the right-invariant 2-form $F$ corresponding to 2-cocycle \eqref{eq:30}
\begin{equation}
\label{eq:32}
F = \mu_1\, dx_1 \wedge dx_2 + (\mu_1 x_1 + \mu_3)\, dx_1 \wedge dx_3 + (\mu_1 x_2 + \mu_4)\, dx_2 \wedge dx_3 + \mu_2\, dx_3 \wedge dx_4.
\end{equation}
It is easy to verify that the conditions $dF = 0$ and ${\cal L}_{\xi_a} F = 0$ are satisfied, where the vector fields $\xi_a$ are defined by \eqref{eq:31a}.

We find the vector potential of the electromagnetic field with 2-form \eqref{eq:32}. For this, we consider the one-dimensional central extension $\tilde{\g} = \g \oplus_{\F} \mathbb{R}$ of the Lie algebra $\g$ corresponding to 2-cocycle \eqref{eq:30}. Let $e_0$ be a basis vector in the one-dimensional ideal $\mathbb{R}$. In this case, the commutation relations of $\tilde{\g}$ in the basis $\{e_0, e_1, e_2, e_3, e_4 \}$ become
\begin{equation*}
[e_1, e_2] = \mu_1\, e_0,\quad
[e_1, e_3] = e_2 + \mu_3\, e_0,\quad
[e_2, e_3] = − e_1 + \mu_4\, e_0,\quad
[e_3, e_4] = \mu_2\, e_0.
\end{equation*}
The corresponding connected simply connected Lie group $\tilde{G}$ as a topological space is homeomorphic to $\mathbb{R}^5$, and it is easy to see that $G \simeq \tilde{G}/(\mathbb{R} \times \mathbb{Z}_2)$. In a trivialization $V \subset G$ of the bundle $\tilde{G} \rightarrow G$, we introduce coordinates of the direct product $\{x_0\} \times \{x_i\}$, where $x_0$ is a coordinate in the one-dimensional subgroup corresponding to the ideal $\mathbb{R}$ and $\{x_i\}$ are coordinates of the second kind on $G$ defined in $V$. The right-invariant vector fields $\tilde{\eta}_a$ on $\tilde{G}$ in this coordinate chart have the form
\begin{equation*}
\tilde{\eta}_1 = \eta_1,\quad
\tilde{\eta}_2 = \eta_2 - \mu_1 x_1 \p_0,\quad
\tilde{\eta}_3 = \eta_3 + \left ( \frac{\mu_1}{2} (x_1^2 - x_2^2) - \mu_3 x_1 - \mu_4 x_2 \right ) \p_0,\quad
\tilde{\eta}_4 = \eta_4 - \mu_2 x_3 \p_0.
\end{equation*}
Choosing the function $U = e^{x_0}$ as a real representation of the subgroup $\mathbb{R}$ and using formula \eqref{eq:17}, we
immediately obtain a particular solution of system \eqref{eq:16}
\begin{equation}
\label{eq:33}
\A_1 = 0,\quad
\A_2 = \mu_1 x_1,\quad
\A_3 = \mu_3 x_1 + \mu_4 x_2 + \frac{\mu_1}{2} \left ( x_2^2 - x_1^2 \right ),
\A_4 = \mu_2 x_3.
\end{equation}
Therefore, the vector potential for the electromagnetic field defined by 2-form \eqref{eq:32} can be chosen in the form
\begin{equation*}
A = \mu_1 x_1\, dx_2 + \left ( \frac{\mu_1}{2} (x_1^2 + x_2^2) + \mu_3 x_1 + \mu_4 x_2 \right )dx_3 + \mu_2 x_3\, dx_4.
\end{equation*}
It can be directly verified that $dA$ coincides with 2-form \eqref{eq:32}.

We proceed to the question of the integrability of Eq. \eqref{eq:03} on the Lie group $G = E(2)\times \mathbb{R}$. For an
arbitrary right-invariant metric, the differential operator $\hat{H}^{(\e)}$ is a second-order operator polynomial in $\eta^{(\e)}_a$, where the functions $\A_a$ are defined by formulas \eqref{eq:33}. It follows from Theorem \ref{th:01} and Statement \ref{st:05} that $\hat{H}^{(\e)}$ admits a symmetry operator algebra isomorphic to the Lie algebra $\tilde{\g} \simeq \left ( \mathfrak{e}(2) \oplus_{\F} \mathbb{R} \right ) \oplus \mathbb{R}$. Taking the coordinate representation of left-invariant vector fields \eqref{eq:31a} and formulas \eqref{eq:06} and \eqref{eq:19} into account, we can easily obtain the explicit form of these operators:
\begin{align*}
\xi^{(\e)}_1 =&\ \xi_1 - \i\e \left ( \mu_1 x_2 \cos x_3 + \mu_3 \sin x_3 + \mu_4 \cos x_3 \right),\\
\xi^{(\e)}_2 =&\ \xi_2 - \i\e \left ( \mu_1 x_2 \sin x_3 - \mu_3 \cos x_3 + \mu_4 \sin x_3 \right),\\
\xi^{(\e)}_3 =&\ \xi_3 - \i\e \mu_2 x_4,\quad \xi^{(\e)}_4 = \xi_4.
\end{align*}

The cohomological index of the Lie algebra $\g$ in our case is
\begin{equation*}
\mathrm{ind}_{[\F]}\, \g = 
\begin{cases}
0, & \text{ если }\ \mu_1 \mu_2 \neq 0,\\
2, & \text{ если }\ \mu_1 \mu_2 = 0.
\end{cases}
\end{equation*}
According to Theorem \ref{th:02}, the KGF equation in an arbitrary right-invariant electromagnetic field on the group $E(2)\times \mathbb{R}$ is integrable if the parameters of 2-cocycle \eqref{eq:30} satisfy the condition $\mu_1 \mu_2 = 0$. The
coadjoint orbit dimension in $\tilde{\g}$ is equal to 2 in this case.

We restrict our consideration to the case where $\mu_1 \neq 0$. Up to the action of automorphisms of $\g$, we can take 2-cocycle \eqref{eq:30} equal to $\F = e^1 \wedge e^2$, which corresponds to the equalities $\mu_1 = 1$ and $\mu_2 = \mu_3 = \mu_4 = 0$.
We consider precisely this case in what follows.

Coadjoint orbits in $\tilde{\g}^*$ are two-dimensional elliptic paraboloids that are level sets of the Casimir functions:
\begin{equation*}
K_0 = f_0,\quad
K_1 = f_1^2 + f_2^2 - 2 f_0 f_3,\quad
K_2 = f_4,
\end{equation*}
where ${f_{\bar{a}}}$ are the linear coordinates of the functional $f \in \tilde{\g}^*$ in the basis $\{e^{\bar{a}}\}$, $\bar{a} = 0, \dots, 4$. We introduce a parameterization of nondegenerate orbits assuming that each orbit in $\tilde{\g}$ passes through the linear functional $\lambda(J) \oplus \e$, where $\lambda(J) = J_1 e^3 + J_2 e^4$, $(J_1, J_2) \in \mathbb{R}^2$. The polarization $\tilde{\pl}$ of the functional $\lambda(J) \oplus \e$ is complex: $\tilde{\pl} = \{e_0, e_1 + \i e_2, e_3, e_4 \}$. It can be shown that there are no real polarizations for this example (see \cite{Dix77}).

We write the $\lambda$-representation of $\tilde{\g}$ corresponding to the chosen polarization:
\begin{equation}
\label{eq:34}
\l_1 = \p_q,\quad
\l_2 = \i (\p_q + \e q),\quad
\l_3 = \i \left ( q \p_q + \frac{\e}{2}\, q^2 + J_1\right ),\quad
\l_4 = \i J_2.
\end{equation}
Instead of the complex coordinate $q$, we can consider a pair of real coordinates $(u, v)$ such that $q = u + \i v$, where $u, v \in \mathbb{R}$. The operators $\l_a$ are skew-Hermitian with respect to the measure
\begin{equation*}
d\mu(q) = e^{\e(q - \bar{q})^2/4} \, \frac{d \bar{q} \wedge dq}{2\i} = e^{- \e v^2} du dv.
\end{equation*}
A delta function on a mixed manifold $Q$ is defined by
\begin{equation*}
\delta(q,\bar{q}') = \frac{\e}{2\pi}\, e^{-\e (q - \bar{q}')^2/4},
\end{equation*}
whence we obtain
\begin{multline}
\label{eq:35}
{\cal D}^{\lambda,\e}_{q,\bar{q}'}(x) = 
\\
\frac{\e}{2\pi} \exp \left [ - \frac{\e}{4} \left ( \bar{q}'^2 - 2 e^{\i x_3} \bar{q}' (x_1 + \i x_2 + q) + (x_1 - \i x_2 + q)^2 + 2 x_2^2 \right ) + \i J_1 x_3 + \i J_2 x_4 \right ]
\end{multline}
by solving system \eqref{eq:23} with "initial" condition \eqref{eq:24}.

We note that it follows from the condition of single-valuedness of function \eqref{eq:35} on the entire group manifold that the real parameter $J_1$ can take only a discrete series of values: $J_1 = 0, \pm 1, \pm 2, \dots$ (the coordinate $x_3$ is periodic). Without going into details, we note that this condition is a restriction on the choice of orbits in $\tilde{\g}^*$ and is called the \textit{integer-valuedness condition} for orbits \cite{Shi00}.

The properties of orthogonality \eqref{eq:25} and completeness \eqref{eq:26} for function \eqref{eq:35} are verified directly.  The measure $d\mu(\lambda)$ for integral orbits is defined by
\begin{equation*}
\int (\cdot)\, d\mu(\lambda) = \frac{\e}{(2\pi)^2}\, \sum \limits_{J_1 = - \infty}^{+\infty} \int_{-\infty}^{\infty} (\cdot)\, dJ_2.
\end{equation*}
Hence, for an arbitrary function $\varphi \in L_2(G, d\mu(x))$, we can define the direct and inverse generalized Fourier transformations according to \eqref{eq:27} and \eqref{eq:28}.

For definiteness, on a Lie group, we consider the right-invariant Lorentzian metric given at unity by the diagonal matrix $\G = \mathrm{diag}(-\varepsilon^2,-1,-1, +1)$, $\varepsilon > 1$. By formula \eqref{eq:18}, we obtain
\begin{equation*}
\hat{H}^{(\e)} = - \varepsilon^2 \p_1^2 - (\p_2 - \i\varepsilon x_1 )^2 - \left ( x_2 \p_1 - x_1 \p_2 + \p_3 + \frac{\i\e}{2} (x_1^2 - x_2^2) \right )^2 + \p_4^2
\end{equation*}
for the operator $\hat{H}^{(\e)}$ in this case.

In accordance with formula \eqref{eq:29}, we write the reduced KGF equation in an external electromagnetic field using the explicit form of $\lambda$-representation \eqref{eq:34}
\begin{multline}
\label{eq:36}
(1 - \e + q^2)\, \frac{\p^2 \psi_{\lambda,\e}(q,\bar{q}')}{\p q^2} + q (1 + \e (2 + q^2) + 2 J_1)\,\frac{\p \psi_{\lambda,\e}(q,\bar{q}')}{\p q} + 
\\
+ \left ( \frac{\e^2}{4}\,q^2 + \e (1 + \e +J_1) q^2 + J_1^2 - J_2^2 + \e \right ) \psi_{\lambda,\e}(q,\bar{q}') = - m^2 \psi_{\lambda,\e}(q,\bar{q}').
\end{multline}
We seek a solution of \eqref{eq:36} in the form
\begin{equation}
\label{eq:37}
\psi_{\lambda,\e}(q,\bar{q}') = e^{- \e^2 q^2/4}\, \vartheta_{\lambda,\e}(z; \bar{q}'),
\end{equation}
where $\vartheta_{\lambda,\e}(z; \bar{q}')$ is a function of the new variable $z = q^2/(\varepsilon^2 - 1)$. We introduce the auxiliary notation
\begin{equation}
\label{eq:38}
\gamma = J_1 - \frac{1}{2} + \frac{\e}{2}\, (\varepsilon^2 + 1),\quad
\delta = - \frac{\e^2}{16}\, (\varepsilon^2 - 1)^2,\quad
\eta = \frac{1}{4} \left( J_1^2 - J_2^2 + m^2 - J_1 + \frac{3}{2} \right ).
\end{equation}
Substituting \eqref{eq:37} in \eqref{eq:36} and using notation \eqref{eq:38}, we obtain the equation for $\vartheta_{\lambda,\e}(z; \bar{q}')$
\begin{equation}
\label{eq:39}
\frac{\p^2 \vartheta_{\lambda,\e}(z;\bar{q}')}{\p z^2} + \frac{(3/2 + \gamma) z - 1/2}{z (z - 1)}\, \frac{\p \vartheta_{\lambda,\e}(z;\bar{q}')}{\p z} + \frac{\delta z - (\gamma - 1)/4 + \eta}{z(z - 1)}\,\vartheta_{\lambda,\e}(z;\bar{q}') = 0.
\end{equation}
The general solution of \eqref{eq:39} is
\begin{equation*}
\vartheta_{\lambda,\e}(z;\bar{q}') = C_1(\bar{q}')\, \mathrm{HeunC} \left ( 0, - \frac{1}{2}, \gamma, \delta, \eta, z \right ) + C_2(\bar{q}') \sqrt{z}\, \mathrm{HeunC} \left ( 0, \frac{1}{2}, \gamma, \delta, \eta, z \right ),
\end{equation*}
where $\mathrm{HeunC}(\alpha, \beta, \gamma, \delta, \eta, z)$ is a solution of the \textit{confluent Heun equation} \cite{SlaLay00}. Substituting the found expression in \eqref{eq:37}, we obtain the general solution of reduced equation \eqref{eq:36}.

In conclusion, we note that a solution of the initial KGF equation in an external electromagnetic field is found by computing integral \eqref{eq:28}.

%%%%%%%%%%%%%%%%%%%%%%%%%%%%%%%%%%%%%%%%%%%%%%%%%%%%%%%%%%%%%%%%%%%%%%%%%%%%%%%%%%%%%%%%%%%%

\end{document}